\documentclass{PoS}

\usepackage{slashed}

\newcommand{\gl}[1]{(\ref{#1})}
\newcommand{\gev}{{\hbox{GeV}}}

\title{W+jets, Z+jets, multijets and new physics searches}

\ShortTitle{$W$+jets, $Z$+jets, multijets and new physics searches}

\author{
  \speaker{Christoph Englert}\\
  Institut for Particle Physics Phenomenology, Department of
  Physics,\\Durham University, Durham, United Kingdom, and\\
  Institut f\"ur Theoretische Physik, Universit\"at Heidelberg, Germany\\
  E-mail: \email{christoph.englert@durham.ac.uk}}
\author{Erik Gerwick\\
 II. Physikalisches Institut, Universit\"at G\"ottingen, Germany
}
\author{Tilman Plehn \\
  Institut f\"ur Theoretische Physik, Universit\"at Heidelberg, Germany
}
\author{Peter Schichtel \\
  Institut f\"ur Theoretische Physik, Universit\"at Heidelberg, Germany
}
\author{Steffen Schumann\\
  II. Physikalisches Institut, Universit\"at G\"ottingen, Germany
}

\abstract{$W$+jets, $Z$+jets and QCD multijet production processes at
  hadron colliders are backgrounds to many searches for physics beyond
  the Standard Model which involve leptons and missing energy in the
  final state. We review the current theoretical and experimental
  status of these processes at the LHC. Furthermore, we discuss
  several methods that allow for reliable predictions for these
  processes in the context of new-physics searches.}

\FullConference{The 2011 Europhysics Conference on High Energy Physics, EPS-HEP 2011,\\
		July 21-27, 2011\\
		Grenoble, Rh\^one-Alpes, France}

\begin{document}

\section{$W/Z$+jets and QCD multijet production at hadron colliders}

At hadron colliders we rely on the appearance of leptons, photons or
missing transverse energy as phenomenological probes of spontaneously
broken electroweak sectors with the addition of a viable dark matter
candidate \cite{review}. Searches for new (renormalizable)
interactions at the LHC face two immediate implications from the
phenomenological success of the electroweak Standard Model (SM): New
physics spectra (not including the Higgs) are either heavy compared to
the weak scale ${\cal{O}}(100~\gev)$ and/or they are weakly
coupled. To constrain, rule out, or even verify any realistic scenario
of physics beyond the SM we therefore have to overcome the
phenomenology-dominating SM backgrounds such as $W$+jets, $Z$+jets or
QCD multijet production. Consequently, lots of effort has been devoted
to the detailed investigation of these processes by both the
experimental and the theoretical communities.

Over the past couple of years there has been remarkable progress in
various aspects of $W/Z$+jets and QCD multijet production
phenomenology, ranging from improved perturbative precision all the
way to first measurements with early LHC data. The latter results
allowed both {\sc{Atlas}} and {\sc{Cms}} to establish the electroweak
SM hypothesis at a new energy frontier by performing Monte Carlo
comparisons and Monte Carlo validation \cite{mcvalidate}.

Furthermore, $W$+jets, $Z$+jets and multijets, acting as ``Standard
Model candle processes'', have been used by the experimental
collaborations for various calibration purposes.  Of particular
importance to, e.g., searches for supersymmetry \cite{jetmis} are
measurements of the jet energy scale uncertainty \cite{jetescale}, the
calibration of the missing energy reconstruction \cite{reconstr} and
the determination of the fake-$\slashed{E}_T$ distribution by detector
effects \cite{fakeetm}.  The performance of one of the most versatile
tools to suppress Standard Model backgrounds in Higgs searches, namely
the central jet veto \cite{cjv}, has only recently been studied at the
LHC in QCD multijet final states \cite{atlascjv}.

\begin{figure}[!b]
  \begin{center}
    \includegraphics[height=5.8cm]{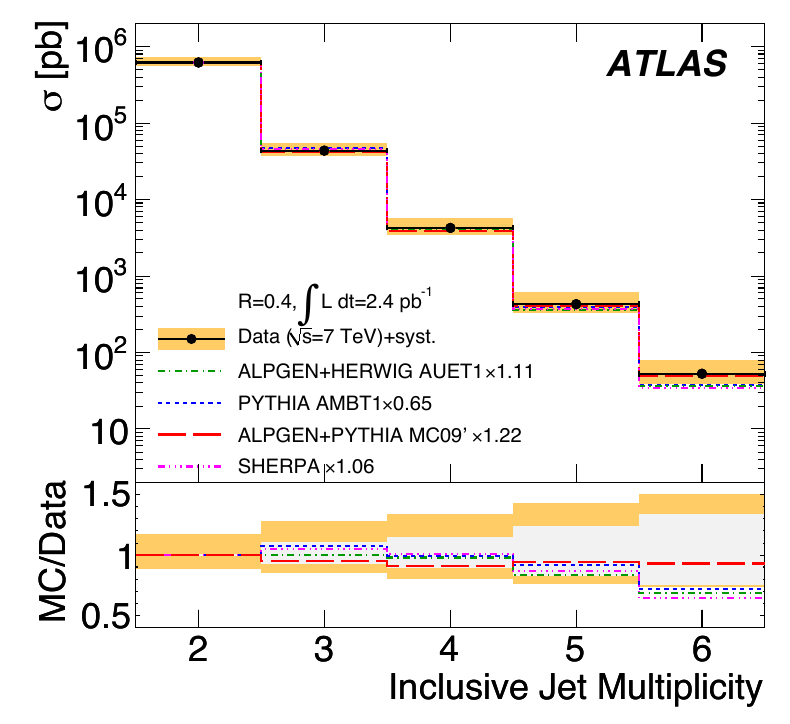}
    \qquad
    \includegraphics[height=5.8cm]{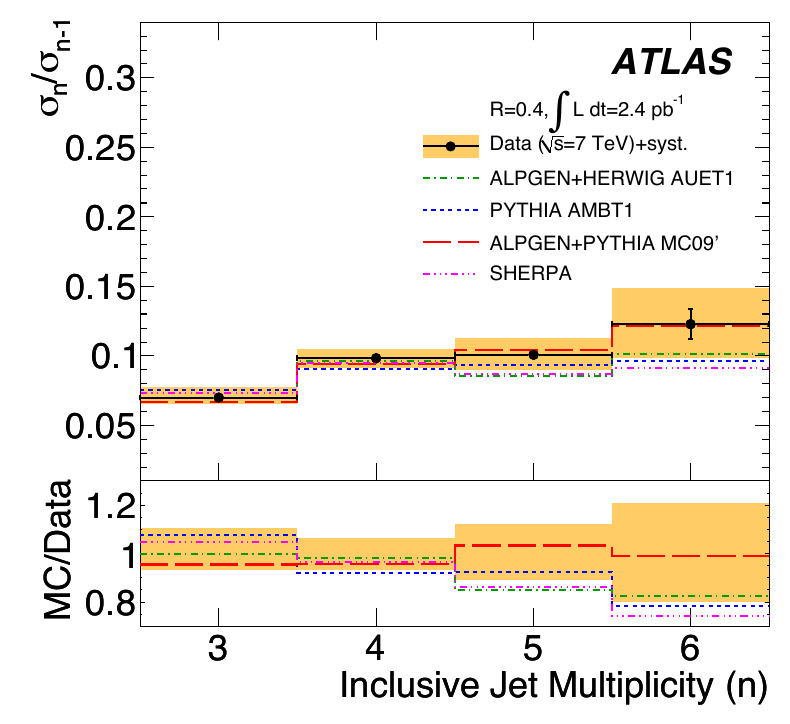}\\
  \end{center}
 \vspace{-0.7cm}
  \caption{\label{fig:qcd} Inclusive QCD multijet production measured
    by {\sc{Atlas}} (figures taken from Ref.~\cite{atlasjets}). The
    impact of the jet multiplicity-correlated pile-up can shift the
    higher $n_{\rm{jets}}$ bins above the MC-expected values.}
\end{figure}

A quantitative knowledge of the impact of QCD corrections on both the
signal and background phenomenology is crucial to connect current and
future LHC results with theoretical predictions. At hadron colliders
such as the LHC, higher order corrections from QCD tend to be large as
a consequence of a sizable amount of initial state radiation. While
early contributions to the field of precise predictions to $W/Z$+jets
and QCD multijet production using Feynman graph-based approaches date
back almost twenty years \cite{Giele:1993dj}, recent developments in
next-to-leading order (NLO) computations involving generalized
unitarity methods allowed the computation of the inclusive production
of $W/Z$ in association with up to four jets \cite{Berger:2009zg}.
Matching these fixed order predictions with parton showers in various
approaches is currently a very active field of research
\cite{Frixione:2010ra}. Due to large contributions from initial state
radiation, matching also provides a good approximation if limited to
the tree level \cite{ckkw} approximation with the overall
normalization obtained from data and/or higher order Monte Carlos.

Progress in multiloop computations has lead to even higher
(next-to-next-to-leading order) precision for selected $2\to 2$
scattering amplitudes, see, e.g., Ref.~\cite{Anastasiou:2001sv}.

\section{$W/Z$+jets and QCD multijet production and new physics searches}

The multijet, multilepton, and missing energy signatures of $W/Z$+jets
and QCD multijets processes are typical signatures that arise in
beyond the SM scenarios with strong interactions and a dark matter
candidate. Disregarding spin correlations etc., the bulk of such
models can be mapped onto the minimal supersymmetric Standard Model on
a phenomenological level. This very popular and well-motivated
extension of the SM was also one of the first new physics
models\footnote{We note that dominant QCD corrections to SUSY
  production processes have become available in a fully automated
  fashion with the {\sc{MadGolem}} package only recently
  \cite{madgolem}.} to be constrained by the LHC experiments using the
jets plus missing energy channel \cite{jetmis} (for a recent update
see Ref.~\cite{jetmis2}).

%%%%%%%%%%%%%%%%%%%%%%%%%%%%%%%%%%%%%%%%%%%%%%%%%
\begin{figure}[!t]
  \vspace{-0.6cm}
  \begin{center}
    \includegraphics[height=6.0cm]{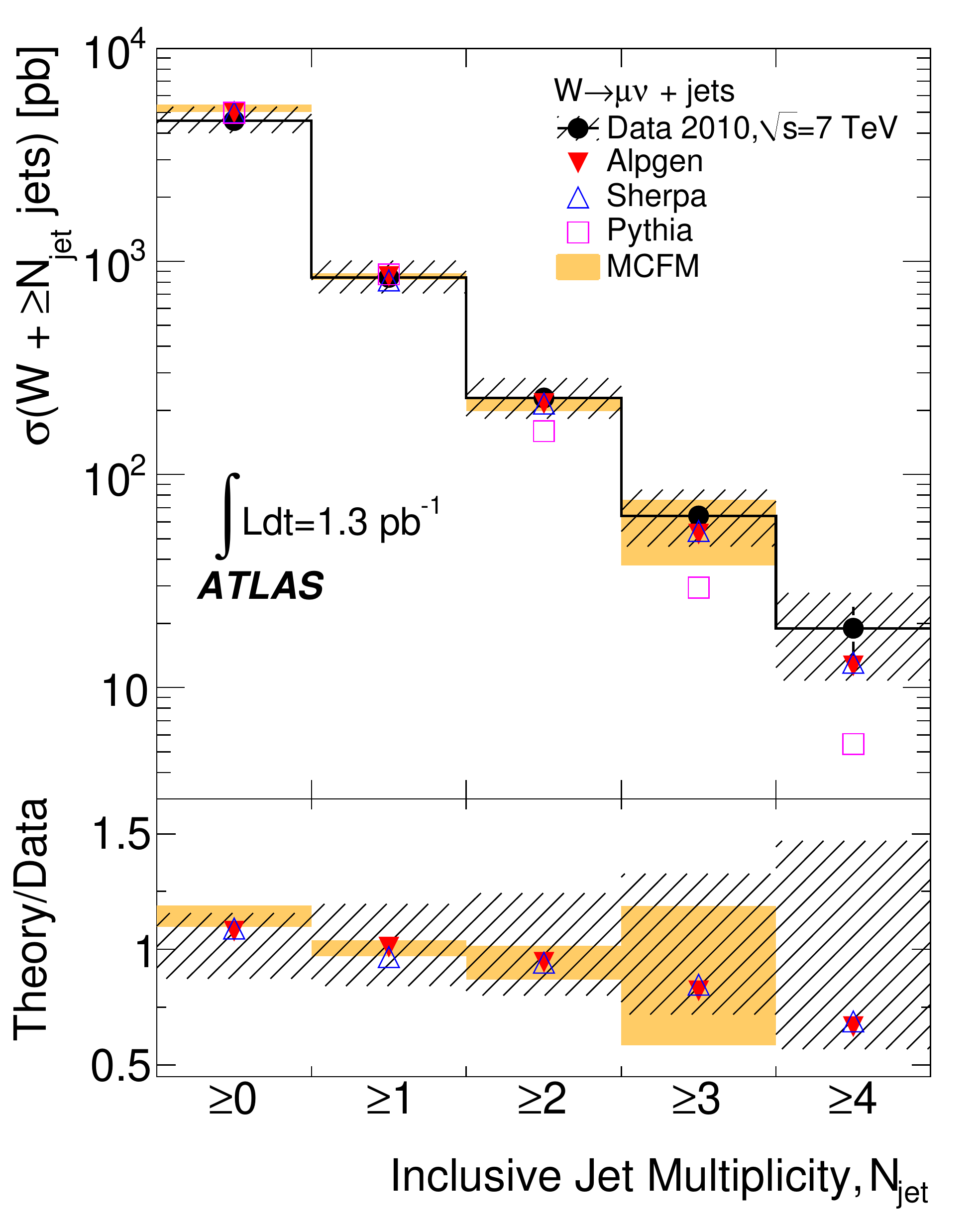}
    \qquad\quad
    \includegraphics[height=5.5cm]{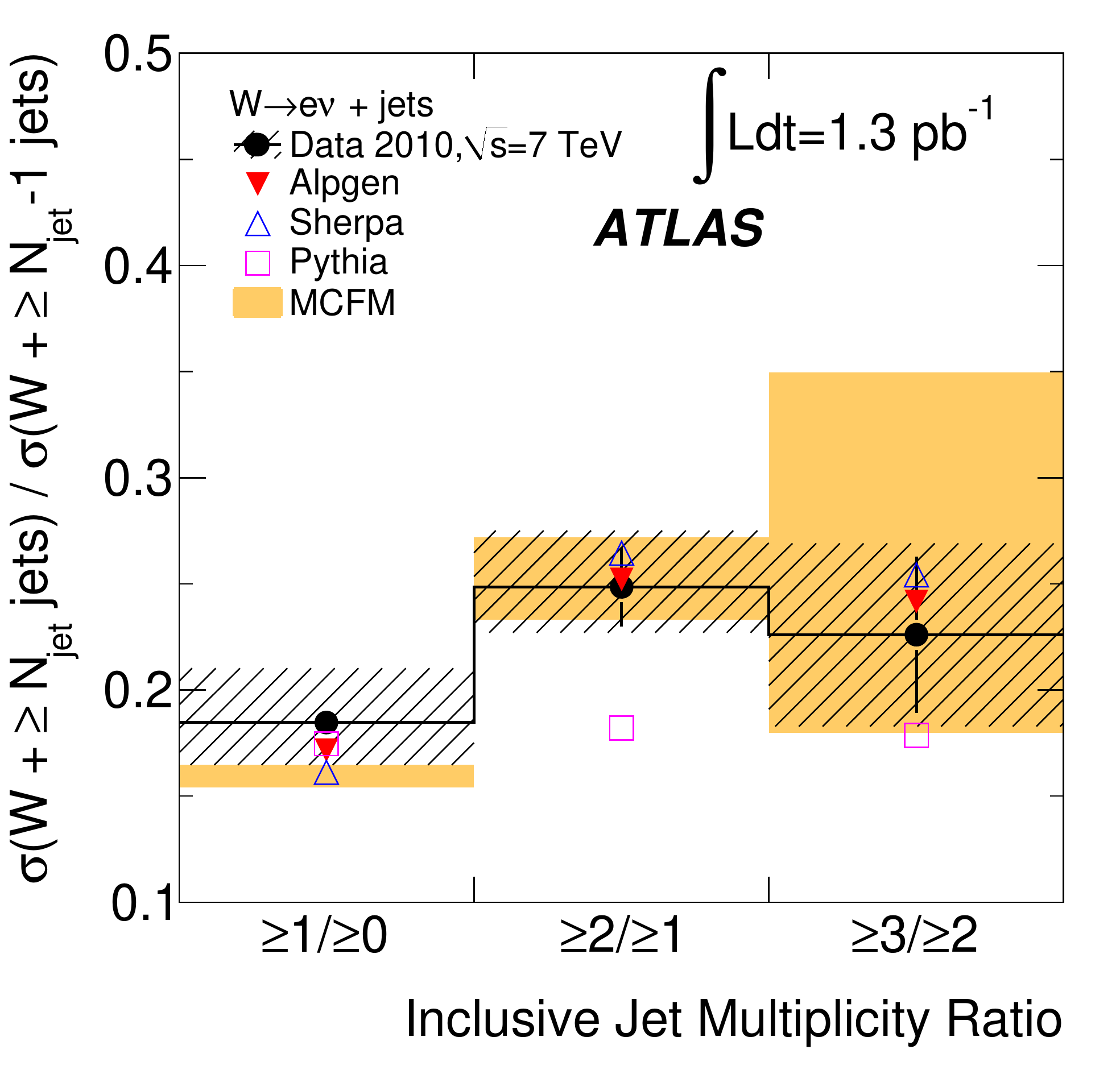}
    \vspace{-0.3cm}
    \caption{\label{fig:wjets} Inclusive $W$+jets production measured
      by {\sc{Atlas}} (figures taken from Ref.~\cite{mcvalidate}).}
  \end{center}
  \vspace{-0.7cm}
\end{figure}
%%%%%%%%%%%%%%%%%%%%%%%%%%%%%%%%%%%%%%%%%%%%%%%%%

While these first results are based on very inclusive cuts and
counting experiments with very small statistics, following the
{\sc{Atlas}} and {\sc{Cms}}~\cite{notes} documentations we expect more
specific analyses to appear soon.  The reason is that in their current
form the analyses can and should be optimized for specific new physics
mass spectra.

\subsection{Model-independent searches in the jets plus missing energy channel}

\begin{figure}
\vspace{-0.6cm}
\begin{center}
\includegraphics[height=5cm]{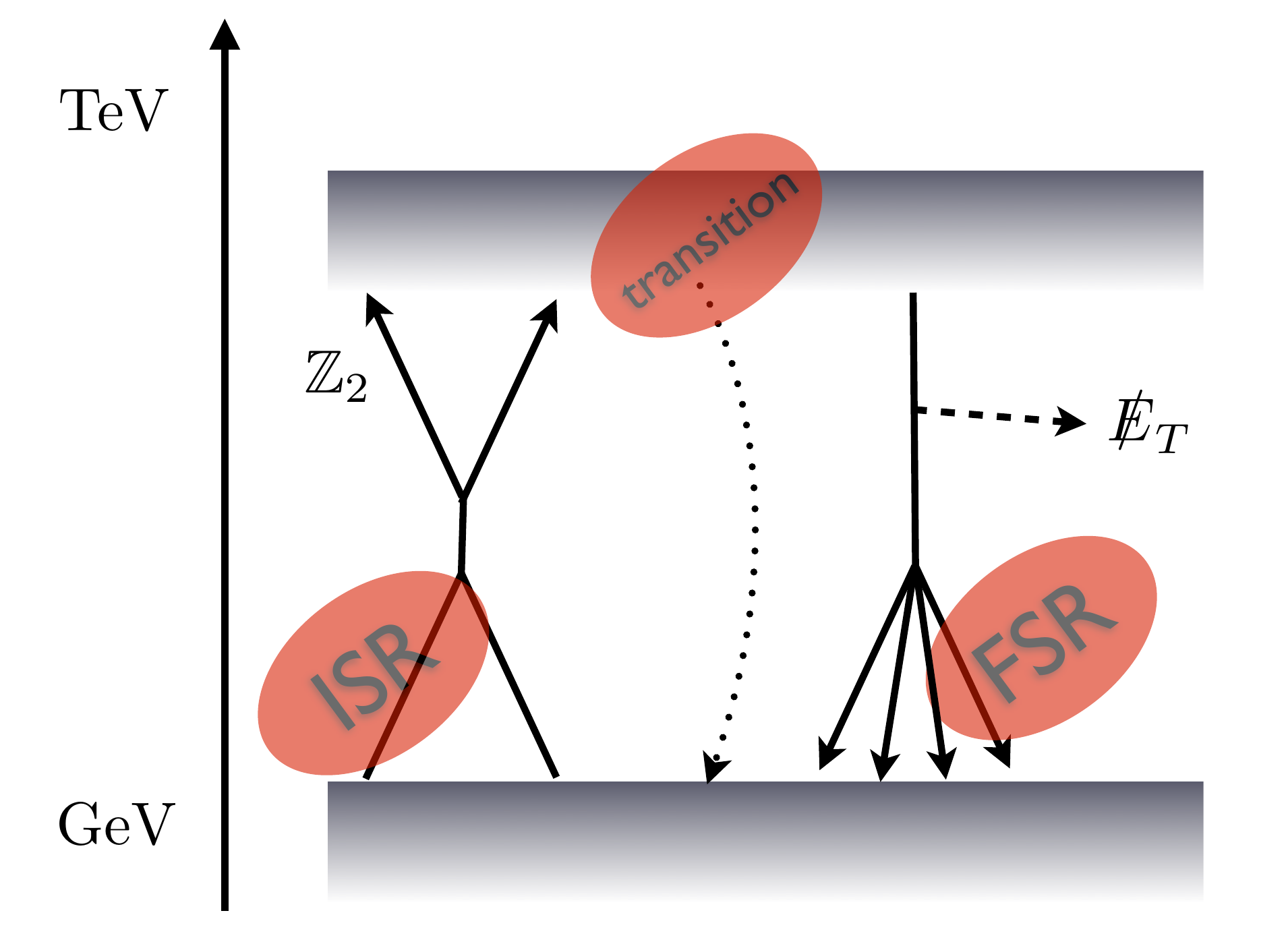}
\qquad
\includegraphics[height=5cm]{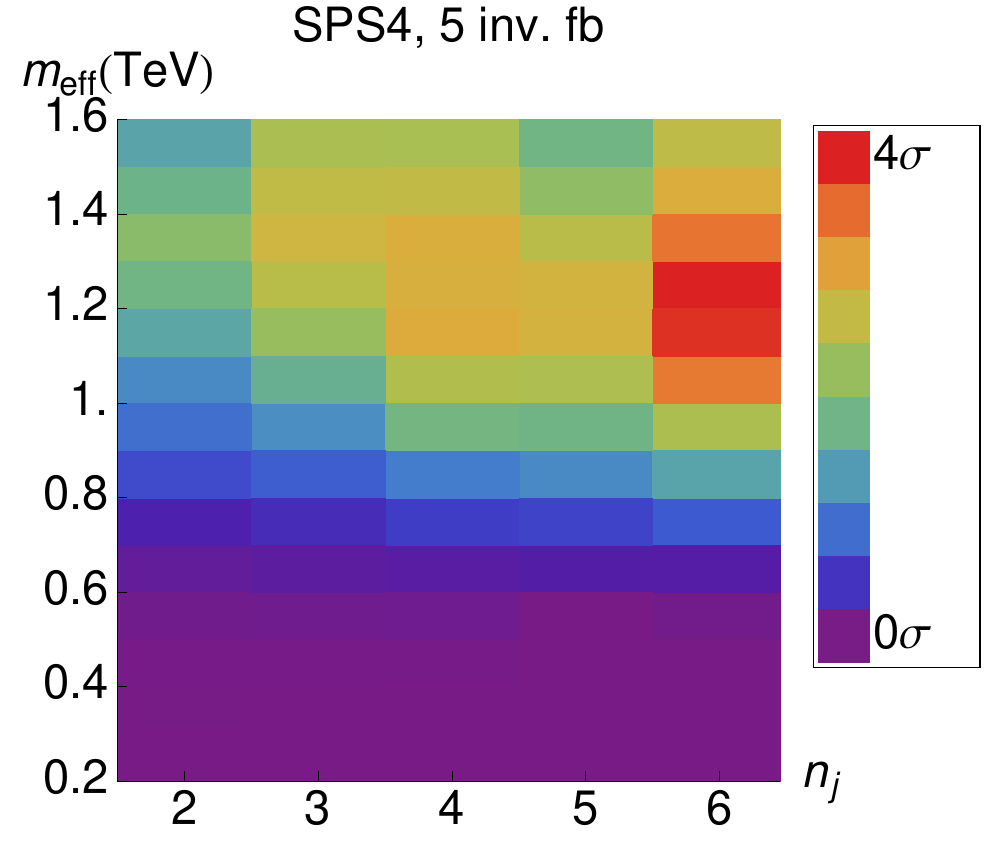}
\vspace{-0.3cm}
\caption{\label{fig:newphysics} Left panel: sketch of a typical
  strongly interacting massive new physics spectrum including a dark
  matter candidate (protected by a ${\mathbb{Z}}_2$) symmetry. Right
  panel: Example of the Likelihood map based on
  $(n_{\rm{jets}},m_{\rm{eff}})$ of SPS4 \cite{autofocus} for a
  luminosity of ${\cal{L}}=5~{\rm{fb}}^{-1}$ and
  $\sqrt{s}=7~{\rm{TeV}}$. For details see text.}
  \vspace{-0.7cm}
\end{center}
\end{figure}

Quite generically, theories of strong interactions which pose a
solution to the WIMP miracle can be pictured as in
Fig.~\ref{fig:newphysics}: Producing pairs of massive new colored
states results in a sizable amount of initial and final state
radiation (ISR/FSR) and a number of hard decay
jets. Depending on the details of the spectrum there might also be
resolvable transition radiation (for a model-independent
phenomenological classification approach see Ref.~\cite{jay}).

%%%%%%%%%%%%%%%%%%%%%%%%%%%%%%%%%%%%%%%%%%%%%%%%%
%%%%%%%%%%%%%%%%%%%%%%%%%%%%%%%%%%%%%%%%%%%%%%%%%

Given that the number of jets distribution of hard jets from QCD and
$W/Z$+jets shows a so-called staircase behavior
(Figs.~\ref{fig:qcd},~\ref{fig:wjets}), i.e. the ratios of inclusive
multiplicities within theoretical and experimental uncertainties is
constant
\vspace{-0.1cm}
\begin{equation}
\label{eq:inclstair}
\hbox{const}=R={\sigma_{n+1}\over \sigma_{n}} \equiv R_{(n+1)/n}
\vspace{-0.1cm}
\end{equation}
for the first couple of bins\footnote{Note that the 3/2 $n_{\rm{jet}}$
  bin ratio (the 2/1 ratio in case of $W/Z$+jets) is notorious because
  of the definition of the underlying hard process \cite{autofocus},
  e.g. jet cuts are trivially fulfilled for a two (one) jet final
  state.}  before phase space suppression causes departure, we can
turn the specific radiation pattern motivated by
Fig.~\ref{fig:newphysics} into an inclusive search strategy
\cite{autofocus} in the jets plus missing energy channel. For $W$+jets
production \cite{Berger:2009zg} it has been shown that the QCD
corrections stabilize the staircase scaling
Eq.~\gl{eq:inclstair}. Most notably, the scaling property of
Eq.~\gl{eq:inclstair} implies the identical behavior for the
{\emph{exclusive}} number of jets via the geometrical series. This not
only opens up the possibility to straightforwardly benefit from the
merits of QCD corrections in an exclusive final state notion, but also
allows to consistently reduce the uncertainties of any other jet
inclusive observable in a data-driven approach. Thereby measuring low
multiplicity bins can be used to constrain the higher ones.

Furthermore, we can utilize the departure from the exclusive staircase
scaling to gain information on a measured resonance's color charge and
mass \cite{autofocus,plehntait}. As an example we show in
Fig.~\gl{fig:newphysics} the result of such an analysis of the SPS4
benchmark point for an integrated luminosity of $5~{\rm{fb}}^{-1}$ at
the LHC with 7 TeV center-of-mass energy. SPS4 has an ``inverted''
mass hierarchy $m_{\tilde q}\sim 750~\gev > m_{\tilde{g}} \sim
730~\gev$ and long decay chains for gluinos through bottom squarks
appear in the high $n_{\rm{jets}}$ bins only. The sensitivity that is
computed from the departure of the exclusive number of jets in a
log-likelihood ratio approach is augmented by a mass scale of the
process, which contains orthogonal information on the new physics mass
scale. The precise definition of the additionally introduced mass
scale to the binned log-likelihood depends on the phenomenological
question we would like to address. To close in on the mass scale of the
SUSY particles we choose the effective mass
$m_{\rm{eff}} = \slashed{E}_T + \sum_{\rm{all~jets}} p_{T,j} \,,$
where the number of jets is again to be understood as an exclusive
quantity.

The described procedure automatically reveals phase space regions that
are inconsistent with the background-only hypothesis. Since the
analysis is as inclusive as possible, sculpting of either the signal or
the background distribution is avoided to large extent and the
inclusive corrections from QCD can be trusted.

\subsection{Predicting $Z$+jets backgrounds to new physics searches}

%%%%%%%%%%%%%%%%%%%%%%%%%%%%%%%%%%%%%%%%%%%%%%%%%
\begin{figure}[t]
\vspace{-0.6cm}
\begin{center}
\includegraphics[height=5.8cm]{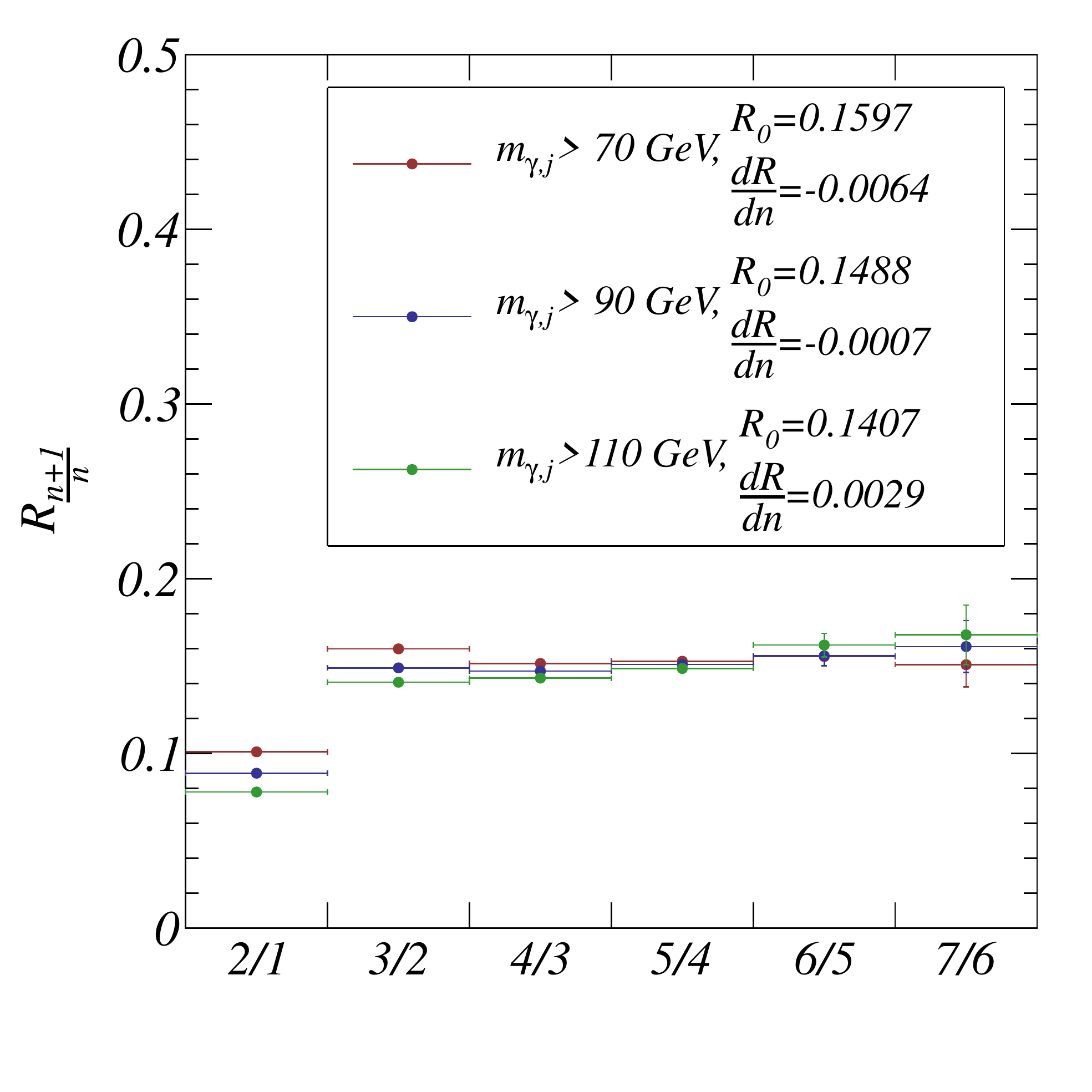}
\qquad
\includegraphics[height=5.8cm]{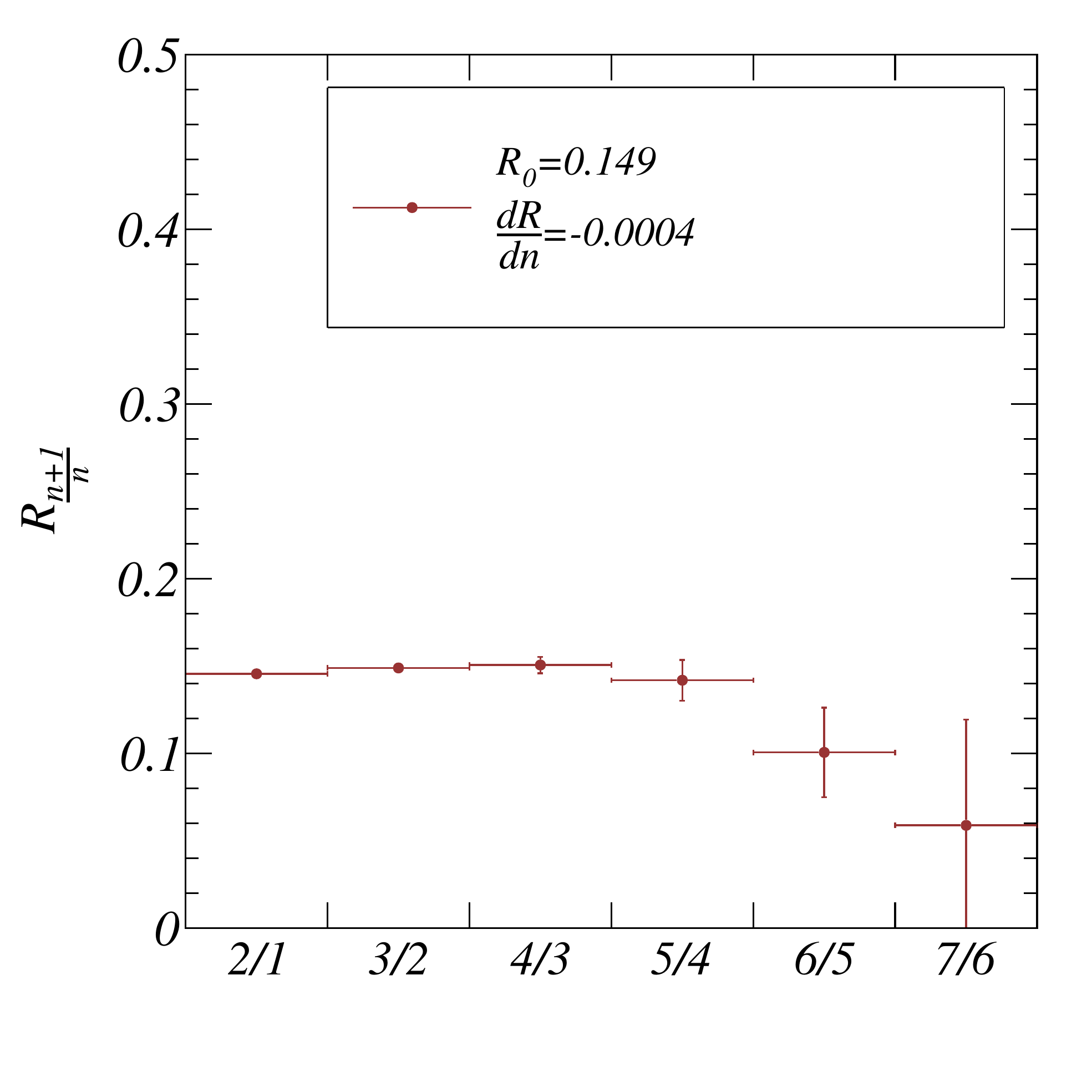}
\end{center}
\vspace{-0.9cm}
\caption{\label{fig:staircase} Staircase scaling in $\gamma$+jets
  (left panel) in comparison to $Z$+jets production (right panel),
  taken from Ref.~\cite{Pjets}. The center of mass energy is
  $\sqrt{s}=7$ TeV.}
  \vspace{-0.3cm}
\end{figure}
%%%%%%%%%%%%%%%%%%%%%%%%%%%%%%%%%%%%%%%%%%%%%%%%%

Another important role is played by $Z$+jets in searches for
supersymmetry, because, with the $Z$ decaying to two neutrinos, it
gives rise to an irreducible background. A strategy which is typically
pursued by the experiments is to extract this background from a
measurement of photon+jets by establishing a phenomenological
translation of $\gamma$+jets production into $Z$+jets
\cite{jetmis}. The knowledge of QCD corrections and theoretical
uncertainties is indispensable to judge on the validity and quality of
such an extrapolation. This issue has been elaborated on recently in
Refs.~\cite{Bern:2011pa,Ask}.

Another question, which is reasonable to ask in the light of the
previous section, is how we can relate a potentially observed
staircase scaling pattern in photon+jets production to $Z$+jets. This
question has been addressed in Ref.~\cite{Pjets}. Due do the massless
photon and its special role in jet fragmentation, observing staircase
scaling in $\gamma$+jets production is non-trivial due to collinear
enhanced phase space regions. Once these are separated off by invoking
a cut on, e.g., the invariant jet-photon mass $m_{j\gamma}\gtrsim
m_Z$, photon+jets production exhibits the staircase pattern in the
exclusive number of jets, which perfectly relates to $Z$+jets. In
Fig.~\ref{fig:staircase} we show the exclusive jet multiplicities
along with the central values of a fit
\begin{equation}
R_{(n+1)/n}^{(staircase)}= R_0 + {dR\over dn} n.
\end{equation}
The size of the error bars corresponds to the limited Monte Carlo
statistics of $10^8$ CKKW-matched events, generated with {\sc{Sherpa}}
\cite{Gleisberg:2008ta}. This opens up an until now unconsidered
observable to further constrain $(Z\to \hbox{invisible})$+jets by
measuring $\gamma$+jets at the LHC.

\subsection{Veto probabilities in Higgs searches}

\begin{figure}[t]
\vspace{-0.5cm}
\begin{center}
  \includegraphics[width=0.47\textwidth]{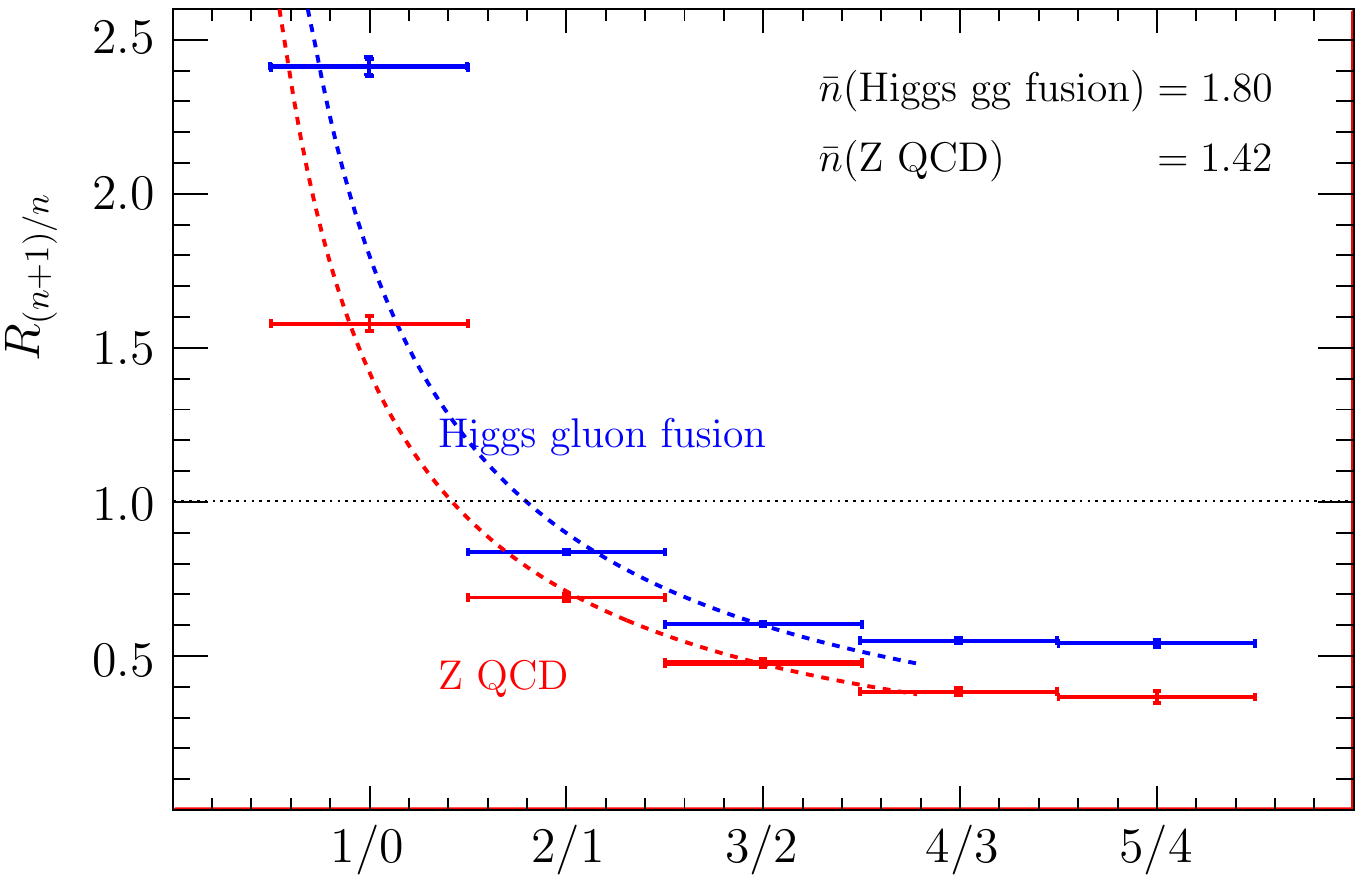}
  \hspace*{.2cm}
  \includegraphics[width=0.47\textwidth]{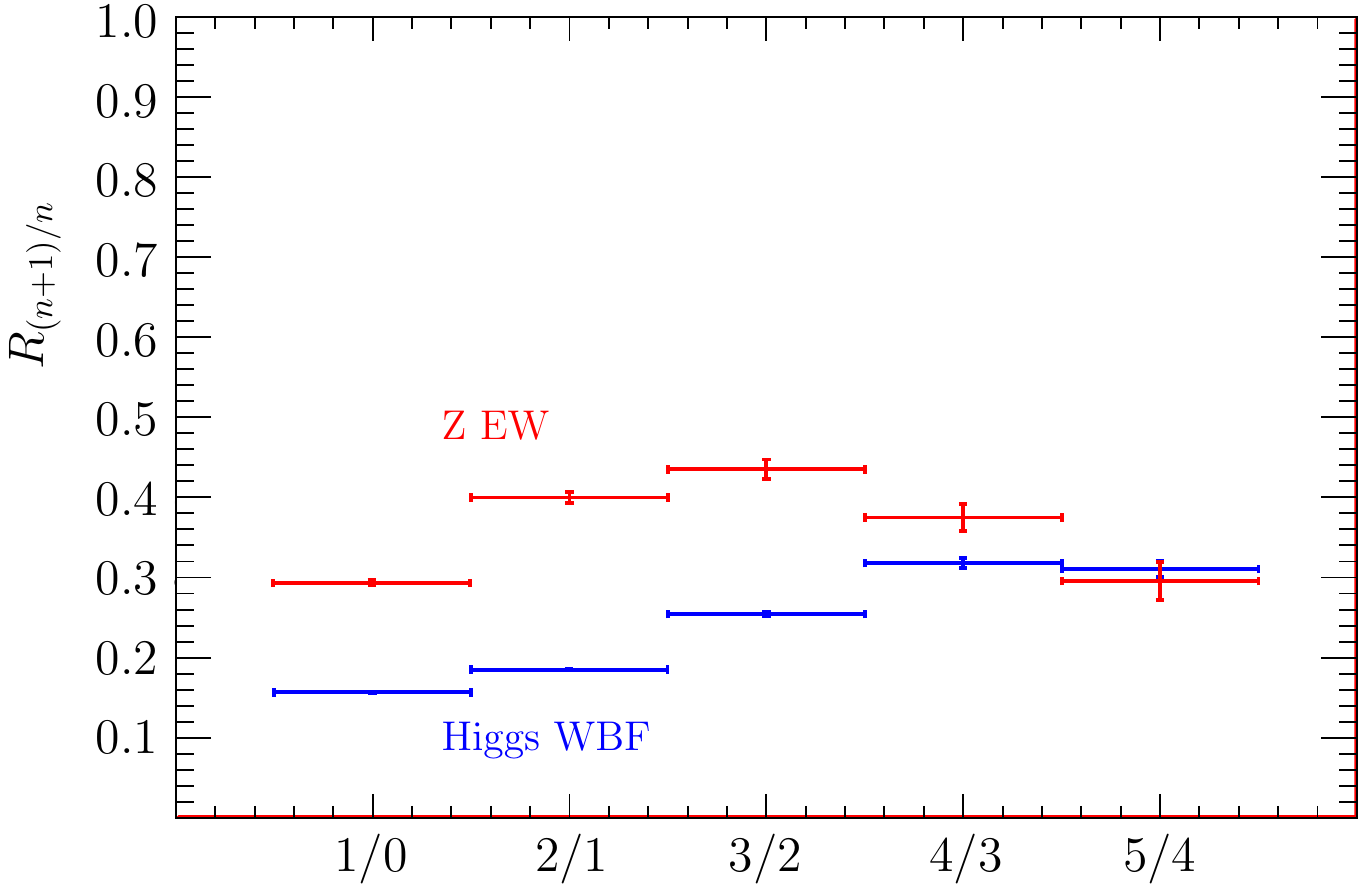}
\end{center}
\vspace{-0.3cm}
\caption{\label{fig:WBF} Exclusive gap jet ratios after WBF cuts:
  (left) Higgs gluon fusion (blue) and Z+jets (Z QCD) at
  $\mathcal{O}(\alpha_s^2\alpha_{EW})$ (red) and (right) WBF (blue)
  and Z-jets at $\mathcal{O}(\alpha_{EW}^3)$ (red).  For the left-hand
  plots the parameter $\bar{n}$ is extracted from the ratio-fit. Both
  figures are taken from Ref.~\cite{higgscale}.}
\vspace{-0.4cm}
\end{figure}

A recent application for jet scaling is central jet veto (CJV)
survival probabilities in searches for the Higgs
boson~\cite{higgscale}.  In weak boson fusion (WBF), requiring hard
($m_{j_1j_2} > 600\,\gev$), widely separated ($\Delta\eta_{j_1j_2} >
4.4$) and opposite hemisphere ($\eta_{j_1}\cdot \eta_{j_2} < 0$)
tagging jets greatly improves signal efficiency \cite{wbf1}.  Reduced
additional central QCD radiation compared with the dominant $Z$+jets
background provides a further distinguishing feature.  A CJV is
therefore imposed on events displaying gap jets with $p_\perp >
20\;\gev$ \cite{cjv1}.  The crucial question is how WBF cuts affect
staircase scaling and thus the extrapolation of the CJV to higher jet
multiplicities. For this purpose a second distinct pattern can be
introduced, \emph{Poisson} scaling, typically associated with
soft-collinear exponentiation. Here the exclusive $n$-jet
cross-section $\sigma_n$ and ratio $R_{(n+1)/n}^{(Poisson)}$ are
defined in terms of the inclusive rate $\hat\sigma_0$ and the expected
number of jets $\bar{n}$ as
\begin{equation}
  \sigma_n =\hat{\sigma}_0 \; \frac{e^{-\bar{n}} \bar{n}^n}{n!} \;\; \quad \quad \;\; 
  R_{(n+1)/n}^{(Poisson)} = \frac{\bar{n}}{n+1}\;.
\end{equation}
Before imposing WBF cuts, both signal and background processes display
staircase scaling as expected.  After cuts, Poisson scaling is
realized in non-color-singlet exchange (see Fig. \ref{fig:WBF}),
notably $Z$+jets at $\mathcal{O}(\alpha_s^2\alpha_{EW})$ and Higgs
production via gluon fusion.  In contrast, Poisson scaling is never
produced for color-singlet mediated processes such as Higgs production
in WBF and $Z$+jets at $\mathcal{O}(\alpha_{EW}^3)$.

%%%%%%%%%%%%%%%%%%%%%%%%%%%%%%%%%%%%%%%%%%%%%%%%%
%%%%%%%%%%%%%%%%%%%%%%%%%%%%%%%%%%%%%%%%%%%%%%%%%

%%%%%%%%%%%%%%%%%%%%%%%%

\end{document}